# Using Cyber Digital Twins for Automated Automotive Cybersecurity Testing

Stefan Marksteiner
*AVL List GmbH*
*Graz, Austria*
stefan.marksteiner@avl.com

Slava Bronfman
*Cybellum Ltd*
*Tel Aviv, Israel*
slava@cybellum.com

Markus Wolf
*AVL List GmbH*
*Graz, Austria*
markus.wolf@avl.com

Eddie Lazebnik
*Cybellum Ltd*
*Tel Aviv, Israel*
eddie@cybellum.com

*Abstract*—Cybersecurity testing of automotive systems has become a practical necessity, with the wide adoption of advanced driving assistance functions and vehicular communications. These functionalities require the integration of information and communication technologies that not only allow for a plethora of on-the-fly configuration abilities, but also provide a huge surface for attacks. Theses circumstances have also been recognized by standardization and regulation bodies, making the need for not only proper cybersecurity engineering but also proving the effectiveness of security measures by verification and validation through testing also a formal necessity. In order to keep pace with the rapidly growing demand of neutral-party security testing of vehicular systems, novel approaches are needed. This paper therefore presents a methodology to create and execute cybersecurity test cases on the fly in a black box setting by using pattern matching-based binary analysis and translation mechanisms to formal attack descriptions as well as model-checking techniques. The approach is intended to generate meaningful attack vectors on a system with next-to-zero a priori knowledge.

## 1. Introduction

The upcoming UNECE regulation R.155 [1] mandates not only the introduction of a cybersecurity management system (CSMS) and according security measures for automotive systems, but also evidence of their appropriateness and effectiveness, which is to be furnished by testing. The regulation becomes effective in Europe for new models in 2022 and for all new registrations in 2024, making it virtually impossible to sell vehicles without structured cybersecurity engineering to the European (as well as the Japanese and Korean) market in the very near future. While the regulation and the underlying security standard ISO/SAE 21434 [2] do not elaborately specify how to test vehicular systems, it is evident that an automated, comprehensive, efficient and scalable automotive cybersecurity testing solution is needed. Due to the characteristics of a common automotive supply chain, which involves many sub-suppliers delivering an original equipment manufacturer (OEM) with heterogeneous, proprietary software with non-disclosed source, this solution to be capable of black box testing, even more as also regulators and other third parties will have an interest in security conformance testing. In order to industrialize automotive cybersecurity testing, this paper outlines an approach that combines an automatic dynamic black box security analysis of a System-under-Test (SUT) with an automated test execution. The requirements for a technical solution to industrialize cybersecurity testing are therefore the capability to a) generate test cases in a black box, automated manner and b) to automate test case execution as much as possible. This is valuable for external testers, regulators and certification bodies to test complete systems, as well as for OEMs and TIER1-x suppliers to verify the claims of their suppliers on subsystems that come to them as black box components. The remainder of this Section outlines preceding as well as distinct work (1.1) and highlights the additional contributions by this paper (1.2). Section 2 a static approach to transfer known attacks from one automotive system to another using an own domain specific language (DSL) and a test case generation producing JSON-based execution instructions is described, while Section 3 describes a system that uses this code in an execution engine to perform the actual tests on an SUT. As this static approach (called *Automated Automotive Cybersecurity Testing - AACT*) requires much a-priori information, we also discuss the concept of a *Cyber Digital Twin (CDT)* for dynamic model creation and data and control flow representation generation that serves as a basis for security analysis, as well as approaches to perform the latter (see Section 4). Three approaches for the synthesis of the dynamic analysis

This research has received funding from the program "ICT of the Future" of the Austrian Research Promotion Agency (FFG) and the Austrian Ministry for Transport, Innovation and Technology under grant agreement No. 867558 (project TRUSTED) and within the the ECSEL Joint Undertaking (JU) under grant agreement No. 876038 (project InSecTT). The JU receives support from the European Union's Horizon 2020 research and innovation programme and Austria, Sweden, Spain, Italy, France, Portugal, Ireland, Finland, Slovenia, Poland, Netherlands, Turkey. The document reflects the author's view only and the Commission is not responsible for any use that may be made of the information it contains.
NB: appendices, if any, did not benefit from peer review.A preprint of this paper has been deposited on ArXiv.

and the automated testing are outlined in Section 5, while Section 6 concludes the paper.

### 1.1. Related Work

There are several known works concerning the usage of digital twins for cybersecurity analysis. The work in [3] applies the digital twins concept to the cybersecurity analysis of smart grids by manually modeling the grid infrastructure and test attack vectors from a threat intelligence system on the digital twin improve the grid architecture's cybersecurity. Another paper [4] provides a method to model a system specification and respective tests for industrial control systems under strict budget constraints. Gehrmann & Gunnarsson [5] describe a method that allows for protecting industrial control systems while being accessible for data sharing by creating a digital twin, using direct state replication through active state monitoring, that can be monitored and acts as a shield for the physical twin. Due to some drawbacks of active monitoring, two related works [6], [7] present a passive state replication approach utilizing the specification of the respective cyber-physical system (CPS) to clone. All of the last three works allow for real-time monitoring systems, which is, however, not a required property of a digital twin that serves for security analysis with the goal of creating test cases. Furthermore, all of the work described so far does not have a focus on automotive systems. The authors of [8] present an approach similar to Gehrmann & Gunnarsson's for the usage in an autonomous driving use case by using sensor data as source of data for the digital twin. Veledar et al. [9] describe a method to model a digital twin in an automotive use case, defining its assets and metrics and facilitate risk management and machine learning-based security-related forecasting. However, all of the works described so far are suitable for performing verification and validation activities in a white-box setting only, as they need very detailed information on the physical system for the synthesizing the digital twin. Aichernig et al. [10] provide a methodology to black box learn a finite state machine via abstract automata learning and derive test cases by executing the model symbolically. This method, however, is meant and only feasible for testing SUTs that contain white and black box components, as the white box components imposes restrictions that allow for the symbolic execution to produce sensible test cases. It is therefore not suitable for purely black box systems.

As the requirements for industrialized automotive cybersecurity testing (as outlined in Section 1) are not met by any of these approaches (combining automation of black box test case generation and execution), this paper describes it own methodology that bases on black box generating a digital twin using pattern matching techniques as outlined in [11] (see also Section 4) and using the a security analysis for generating and executing test cases (see Section 5). The groundwork static process for the latter (described in 2) also orients on an automotive cybersecurity testing architecture outlined in [12].

### 1.2. Research Contribution

The presented work contributes and approach for fully automated black box security testing virtually no a priori knowledge using three strategies. Based on existing work for deriving a cyber digital twin and performing security analysis based on pattern matching [11], we discuss to
a) Transform an existing data flow representation in a finite state model and evaluate faults on that model to derive test cases;
b) Find interesting edge cases by performing model checking on that model;
c) Transform the analysis results into generic attack descriptions that be used to generate test cases.

The test case generation and execution

## 2. Static Approach to Automating Automotive Cybersecurity Testing

One of the key issues in security test industrialization is portability, i.e. to be able to transfer a cybersecurity attack (or test case) from one system to another. The reason is:
a) To expand the usage of one engineered attack vector beyond a single system (scalability);
b) To allow for benchmarking different SUTs (comparability);
c) To improve patterns for single steps of a tests and easily re-use that improvement (efficiency);
d) To put the test cases in a defined workflow that needs only minimal user interaction (automation).

In order to fulfill these targets, our methodology is to abstract a concrete test case and turn it into a generic test scenario by stripping it of all SUT-specific information. Single executable steps of a test case (test scripts) become generic test patterns [13]. At the test case generation, the abstract test scenario is concretized using information from an SUT database (see Figure 1), generating a test case out of a scenario.

For modeling and storing these generic attacks, we developed a domain-specific language (DSL), called *Agnostic Language for Implementing Attacks (ALIA)* [14]. For (a simplified) example, an attack that captures an infotainment head unit and issues a fake speed signal onto a connected CAN bus would not contain any specifics of the SUT, rather a CAN message for the speed signal would generically called CAN_SPD, while a test case generation fuses the script with information about the SUT (in this exemplary case, the concrete CAN message, e.g. 5A1#11.2233.44556677.88[1]). Listing 1 shows an example DSL attack script, where the *Actions* part contains the actual attack (while the *PreConditions* define when to omit a step ant the *Postconditions* contain information for the test evaluation): First can for a potential victim to a *BlueBorne* attack (line 5), then exploit a found target to get access (line 6), install a malicious script (line 7 - in this case a DoS on the CAN bus) and execute the attack (line 8) by using the script from the line above. The outcome is a semi-executable JSON script that will be interpreted and executed by a dedicated execution engine (see next Section).

---

1. This is just an example. In this case, 5A1 is the object identifier that determines the message content (e.g. 'brake', 'RPM', 'steering torque'), while the rest is the message content. Which ID belongs to which function, as well as the meaning of set bits is proprietary and defined solely by the manufacturer.

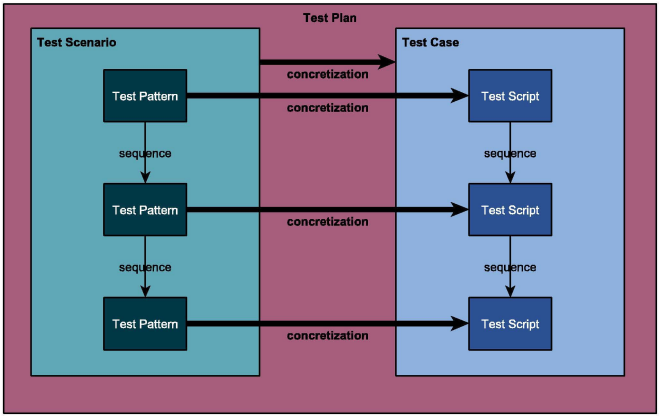

Figure 1. Test Abstraction as in [13]

Listing 1. DSL Attack Script Example from [14]

```
1  PreConditions:
      BT-Scanning: BT_IF
3     BT-Exploiting: target
   Actions:
5     BT-Scanning: target = scan(type:BlueBorne, interface BT_IF)
      BT-Exploiting: shell = exploit(type: Blueborne, target:target)
7     Install Script: attackScript = exploit(type:InstallAndroidCANDosScript
          , target:target)
      can_attack: exploit(type:ScriptExecution, target:target, shell:shell,
          file:attackScript)
9  PostConditions:
      BT-Exploiting: shell
11    can_attack: CAN_MESSAGE(CAN_SPD)
```

## 3. Test Execution

The AACT Test Execution is runs on the *Attack Execution Engine (AXE)*, a platform independent python application, which is based on the Flask framework, a lightweight Web Server Gateway Interface (WSGI) framework that is designed to enable an easy start for web applications but also to allow easy upscaling for complex applications. The core functionality of flask is a wrapper around the Werkzeug framework and the Jinja template engine . For our scenario, the AXE runs on Kali Linux, a Debian-based Linux distribution which is optimized for Security Auditing and Penetration testing. It includes over six hundred tools for penetration testing, security research, computer forensics and reverse engineering, which means that most of the software utilities needed to execute the security cases is already installed out-of-the-box. The hardware could be an ordinary PC or even a Raspberry Pi that posses a direct CAN connection for testing (e.g. a PiCAN2 board for the Raspberry). For Bluetooth connection, a Cambridge Silicon Radio (CSR) USB device (i.e. dongle) is necessary, as a those allow for changing the MAC address arbitrarily. By providing various resources via a restful API, the application takes HTTPS POST requests that contain JSON objects as input via and processes them according to the specified URL path. The JSON data interchange format is a subset of JavaScript and allows transfer data as name/value pairs between applications in an easily readable and writable manner [15]. Input requests for the application contain an array of executable commands, which each consist of the tool to use, its parameters, the environment and a time duration that specifies how long the output collection phase takes. The parameter list for each command may include placeholders that are either determined by the application at runtime or are loaded from the global configuration of the application before execution. The Test Case Generator (TCG) uses scripts that are defined in the attack DSL as a blueprint and outputs corresponding JSON objects that can be directly used for execution and consist of the respective Precondition and Action block, whereas the postcondition block is forwarded to a test oracle. Each step in the Action block is executed subsequently. Before execution, the application checks if all corresponding preconditions are matched. Depending on the necessary execution environment, commands can be executed in different shells than the initial bash shell as well, for example if an exploit returns a reverse shell, it is stored onto an object and new commands can be piped into that shell as an input. After the execution, the output of each command is collected and stored into the HTTPS response of the application. Verification of Postconditions is done by the Test Oracle, which is implemented as a rule-based engine that runs on an existing automotive test control solution. The Oracle receives the condition block from the TCG and monitors the SUT and the tool output received from the AXE accordingly. If a condition is met, it reports this back to the Orchestration Software. Through the rules, it asserts whether the SUT has failed or passed a specific test of the complete test case and reports this result to the orchestration software and GUI.

## 4. Dynamic Digital Twin Generation

A Digital Twin (DT) produces a virtual model of a physical object as a digital representation with the purpose of simulating them before construction to facilitate predictive maintenance [16]. A Cyber Digital Twin (CDT) transfers this idea of the DT to automotive software [11]. As such, a CDT digitally represents the firmware of a vehicular component, e.g. of an electronic control unit (ECU) or a head unit of an infotainment system and can be used for thorough security analysis. In general, most software utilizes widespread software packages to build on. This is particularly true for the automotive industry, where original equipment manufacturers (OEMs) assemble parts of suppliers (TIER 1), which in turn use parts of sub suppliers and so on (TIER 2-X). This applies also for the software in ECUs and other integral parts of an automotive system. Mostly in these settings the source code of the respective firmware is not available, rendering the part in question essentially a black box. For comprehensive security analysis, a system that is capable of extracting an SUT's (e.g. an ECU's) behavior-defining key attributes is necessary. This is given by the CDT approach, that extracts these attributes automatically, which can be used for the analysis and, subsequently, to generate test cases by converting them into ALIA statements or induced faults (see next section) that can be converted into executable attack code. To do so, the firmware of the SUT is automatically transformed into a corresponding CDT to be used for cybersecurity analysis. Due to the circumstances of the automotive supply chain described above and the fact that the automotive domain predominantly works with proprietary, closed-source products, the firmware is usually only available in binary form, which mandates the CDT approach to be operational without access to source code or deeper inside-details of the firmware. The CDT creation engine generates a software bill of materials (BOM) that contains all libraries and components of the SUT by using pattern recognition algorithms that compare software patterns of the SUT with known applications and modules. The CDT engine automatically discovers all available interfaces (e.g., CAN Bus, GPS or Bluetooth), employed software libraries (e.g., *OpenSSL* or *SQLite*) and further information. To create a model that is suitable for security analysis based on dynamic executions, the engine also extracts the control and data flow. Mainly the CDT encompasses the following attributes of the SUT (and the interaction between them):

- Software & Hardware bill of materials (S&H BoM);
- Network interfaces;
- Operating system and the Operating system settings;
- Kernel configuration;
- OS-level security configuration;
- Memory management and mapping;
- User credentials;
- Firewall configuration;
- Application frameworks in use and their configuration;
- Available and in-use APIs;
- Application configuration;
- Encryption mechanisms and flows;
- Encryption keys;
- Control and data flow representation.

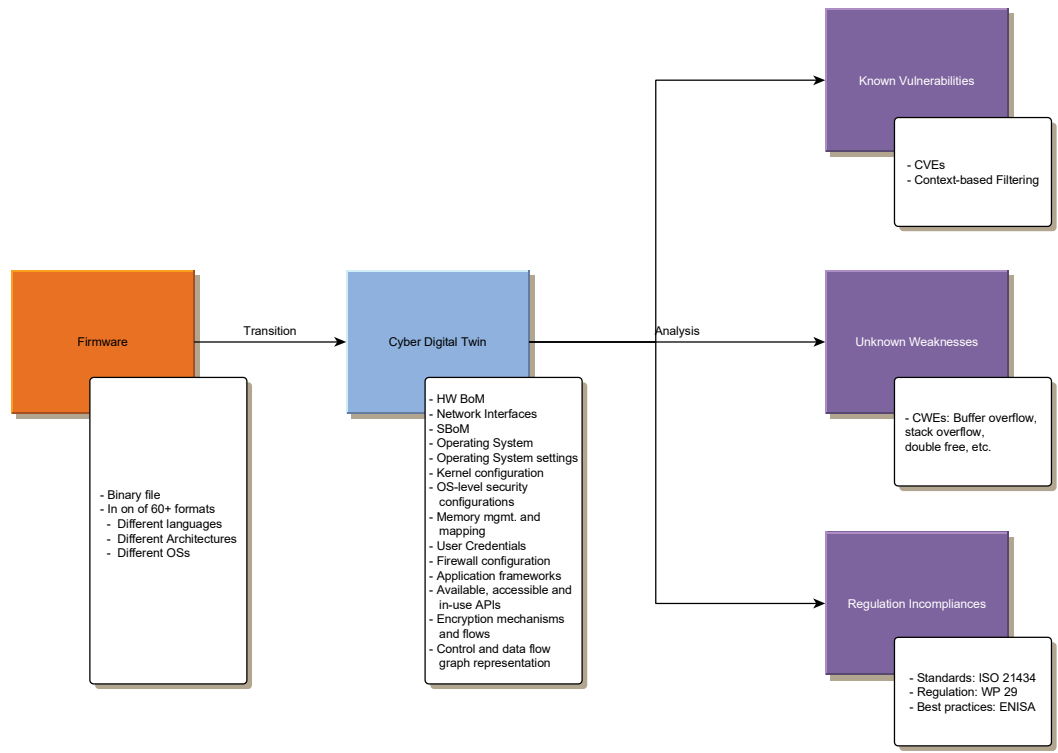

Figure 2. Digital Twin Derivation

Using the very same pattern recognition techniques, the CDT is the analyzed for

a) Known vulnerabilities derived from Common Vulnerabilities and Exposures (CVE) databases;
b) Unknown weaknesses as classified by the Common Weakness enumeration (CWE) scheme;
c) Policies and compliance rules.

Figure 2 gives an overview on the CDT generation and analysis.

## 5. Digital Twin-based Security Testing

In order to transition the model of the Cyber Digital Twin into test cases (that can eventually be

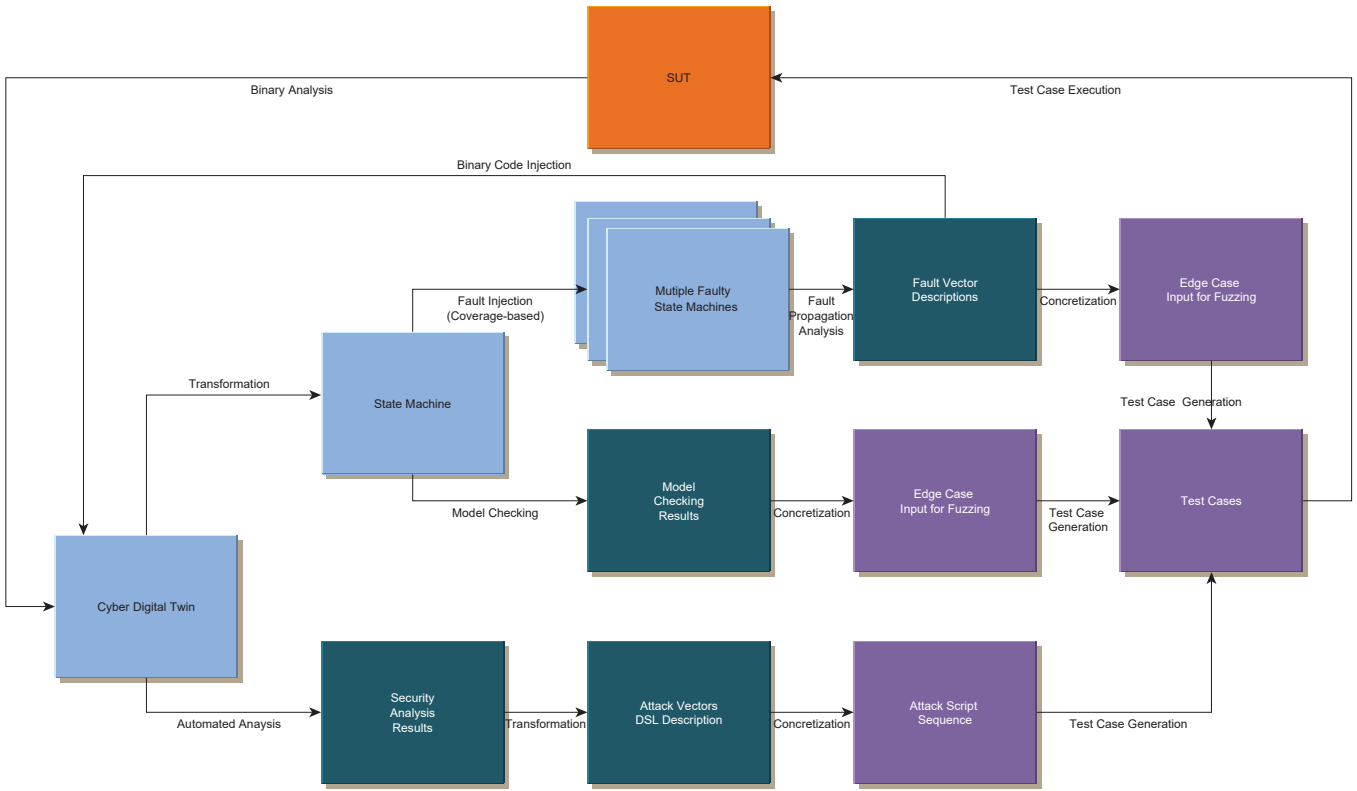

Figure 3. Test Case Generation Strategies

executed on the AXE) we follow two different principle approaches:

- Transforming the CDT security analysis results into attack vectors;
- Transforming the CDT model into a state machine and using this as input fault injection and model checking.

The former approach mandates a mechanical algorithm that transforms the analysis results into a DSL description. This occurs by a) referring to the vulnerabilities identified in the CDT security analysis and mapping proper attack vectors (and later in the toolchain exploits) to it and b) using building block attack vectors and exploits that would, for instance, try to issue a payload into an identified buffer overflow attack. This resembles the static workflow, however, with a dynamically generated starting point. Still, there are building blocks and exploit code necessary to be in place a priori. The second approach uses methods traditionally attributed to automated test case generation and to formal verification, respectively. To fully utilize already established methods, we transform the CDT model into a state machine. This state machine allows for two operations:

- Inject faults into the model using mutation-based algorithms;
- Examine security-relevant parts of the machine through model checking.

The first operation uses state-of-the-art model-based testing methodology. This includes input and fault injection to the model using mutation-based algorithms [17] and potential paths to outside interfaces of the latter using distance heuristics [18].

The second operation includes firstly extracting security-relevant parts, which are identified by the bill of materials (BOM) provided by the CDT. These relevant parts of the model are then checked by a model checker [19]. At those points where the model checking fails, the respective input is taken to form input for directed fuzzing tests [20], [21]. Figure 3 depicts these approaches with blue boxes being models, cyan being checking results, purple being concrete test case parts and orange being the SUT. The result of the binary (pattern matching) analysis therefore serves either as a basis for synthesizing a state machine for model-based testing or as a basis for an agnostic attack description using a DSL.

## 6. Conclusion

Due to regulations and standards, an industrialization of automotive cybersecurity testing is heavily needed. This paper outlined an approach to provide a tool for automated, comprehensive and efficient cybersecurity testing of vehicular systems. The approach displays how test automation is possible by transferring attacks from one system to another through generalization using a DSL, but also how to derive a Cyber Digital Twin model of an SUT that can be analyzed for cybersecurity properties. The model and result of this dynamic, black box model building and analysis can then be transformed into test cases for the automated testing system using fault injection, model checking as well as transformation into the DSL and appropriate attack script selection in the test case generation, as well as automated execution. This way, the complete system is capable of automatically analyzing the cybersecurity of an automotive component and subsequently generating and executing tests to verify its security, which makes this approach most suitable for external testing facilities to test vehicular systems with minimal or no a priori knowledge about the SUT. Future work includes, apart from implementation tasks, mainly methodologies for advanced case generation out of the derived attack descriptions, as well as methods to automating code transfer between SUTs (e.g. address estimation for exploits).